\documentclass[10pt,conference,letterpaper]{IEEEtran}
\usepackage{times,amsmath,epsfig}
\makeatletter
\newif\if@restonecol
\makeatother

\usepackage{algorithm2e}

\usepackage{graphicx}
\usepackage{balance}  % for  \balance command ON LAST PAGE  (only there!)

\title{A Simple and Efficient MapReduce Algorithm for Data Cube Materialization}

\author{%
{Mukund Sundararajan{\small $~^{1}$}, Qiqi Yan{\small $~^{2}$} }%
\vspace{1.6mm}\\
\fontsize{10}{10}\selectfont\itshape
\,Google Research\\
1600 Amphitheatre Pkway, Mountain View, CA, 94043, USA\\
\fontsize{9}{9}\selectfont\ttfamily\upshape
$^{1}$\,mukunds@google.com, $^{2}$\,qiqiyan@google.com%
\vspace{1.2mm}\\
\fontsize{10}{10}\selectfont\rmfamily\itshape
}

\begin{document}

\maketitle

\begin{abstract}
Data cube materialization is a classical database operator introduced in Gray et al.~(Data Mining and Knowledge Discovery, Vol.~1), which is critical for many analysis tasks. Nandi et al.~(Transactions on Knowledge and Data Engineering, Vol.~6) first studied cube materialization for large scale datasets using the MapReduce framework, and proposed a sophisticated modification of a simple broadcast algorithm to handle a dataset with a 216GB cube size within 25 minutes with 2k machines in 2012. We take a different approach, and propose a simple MapReduce algorithm which (1) minimizes the total number of copy-add operations, (2) leverages locality of computation, and (3) balances work evenly across machines. As a result, the algorithm shows excellent performance, and materialized a real dataset with a cube size of 35.0G tuples and 1.75T bytes in 54 minutes, with 0.4k machines in 2014.
\end{abstract}

\section{Cube Materialization}
\label{problem}

As a concrete example, in the context of search engine advertising, a typical data analysis task can involve a dataset like in Table 1. Here for each region where the search engine's users are from, for each category of queries the users entered, and for each advertiser who advertised on the result pages of the queries, we have the total count of ad impressions this advertiser showed.
\begin{table}[tbh]
\label{input-table}
\centering
\caption{Input dataset}
\begin{tabular}{|c|c|c|c|c|c|} \hline
country & state &  city & query- & advertiser & count \\
&& & category & &  \\ \hline
US& CA& Mtn View & Retail & Amazon & 400  \\ \hline
CN& ZJ& Hangzhou & Shopping & Taobao & 300 \\
..&.. &.. & .. & .. & .. \\ \hline
\end{tabular}
\end{table}

In general, in this paper, a dataset can contain a number of discrete hierarchical dimensions, and additive metrics. Here each hierarchical dimension is composed of some number of columns, with higher-level columns appearing to the left. E.g., the region dimension has three columns country, state, and city, and the advertiser dimension has just one column. For simplicity, we will deal with a single metric called count in the paper, with the understanding that our results extend easily to multiple metrics and algebraic measures~\cite{G+97}.

Many analysis tasks are then concerned with the aggregate metrics for subsets of rows that can be defined by specifying concrete values for a subset of the columns, aggregating over the other columns. We call such subsets of rows as \textbf{segments}. (In the case of a hierarchical dimension, if a value is set for a lower level column (such as state), a value must be set for all higher level columns (such as country) as well.)

For example, we could ask for the total ad impression count for the segment defined by country=US, state=*, city=*, query-category=* and advertiser=Amazon, where * means the column is aggregated over all its possible values. The \textbf{key} of a segment is the vector of the form (``US", ``CA", *, ``Retail", ``Amazon"), listing all the columns.

Cube materialization was introduced in Gray et al.~\cite{G+97}, which refers to the task of computing counts for all segments. Cube materialization is an important problem. By making all these segments' counts precomputed and hence instantly available, it enables real-time reporting, efficient online analytical processing, and many data mining tasks~\cite{G+97}.

Note that the number of all segments can be large, because for a dataset with $n$ dimensions, there can be $2^n$ segments that each input row contributes to. We call the ratio of number of outputs to number of inputs the \textbf{blow-up ratio}. The blow-up ratio can be easily greater than 10 for many datasets. Hence we measure the scale of a cube materialization problem by the size of the cube, which refers to the set of all segments.

\section{Three Important Factors for Efficiency}

With the recent surge of ``big data", we need to do cube materialization for larger and larger datasets, and the focus of this paper is to do this at scale with a parallel computation framework such as MapReduce \cite{DG04}. Nandi et al.~\cite{NYBR12} was the first paper to study cube materialization at large scale, for the class of partially-algebraic measures. We focus on additive and algebraic measures, and propose a simpler algorithm with better scaling properties. To aid our discussion, we identify three important factors for efficient algorithms below. Our algorithm was designed in a way that was focused around these factors, and we will compare our algorithm to Nandi et al.'s with respect to these factors.

\textbf{Minimizing Copy-Add Operations / Messages}: In cube materialization, a basic operation or unit of work is copy-add. For example, to compute the aggregate for the segment of country=US, one option is to copy and add the counts of all input rows with country=US onto an aggregating variable for the segment of country=US. We note that all existing algorithms are based on this basic copy-add operation.

\textit{Messages}: We call such an copy-add operation a message, as it can be seen as every input row sends its aggregate as a message to the segment of country=US. A good algorithm should avoid an excessive number of messages.

\textbf{Locality}: A copy-add operation a.k.a., message can be a remote one or a local one, depending on e.g.~whether each input row with country=US is located in a different machine from the count for the segment of country=US. A remote message is well-known to be more expensive than an intra-machine local message, by one or two orders of magnitude, and should be avoided as much as possible.

\textit{Remark}: Since cube materialization often leads to a blow-up in data size, it is often coupled with aggressive output filtering, with most output rows never written out. For this reason, we will ignore the cost of output writing, and do not count them as remote messages.

\textbf{Balance}: To avoid the running time being dominated by a few straggling machines, a good algorithm needs to partition the data and all the local messages to be evenly distributed across machines. (Remote messages are evenly distributed automatically due to random sharding.) The challenge here is to deal with various kinds of data-skewness in real datasets. For example, if we shard the data based on advertiser's id, the sharding can be uneven if a big advertiser contributed a significant fraction of input rows to the dataset.

\section{On Prior Art}

Nandi et al.~\cite{NYBR12} proposed the first MapReduce algorithm for cube materialization. The starting point of their work was the following naive broadcast algorithm.

\begin{algorithm}
\KwSty{Mapper}: For each input row, enumerates all segments that the row belongs to (putting *'s in the columns in all possible ways that respect the hierarchies), and sends the input row's count to every such segment.\\
\KwSty{Reducer}: There is one reducer for every segment with non-empty data, and the reducer simply aggregates the counts from all incoming messages.
\caption{Naive Broadcast}
\end{algorithm}

Nandi et al.'s algorithm then does the following optimizations over the naive broadcast algorithm: 

Regarding message minimization, in the naive broadcast algorithm, an excessive amount of messages are sent for each input row ($2^n - 1$ if there are $n$ one-column dimensions). Nandi et al.'s algorithm then uses heuristics to batch different cube regions (sets of segments specified by which columns are aggregated) to ``batch areas" so that each input row only sends to every batch area. This alleviates the problem to some extent. However, the construction of batch areas must obey certain validity constraints, and the number of batch areas can still be high. It is 28 for the largest dataset mentioned in Nandi et al., and probably higher for an even larger dataset like ours with more dimensions.

Regarding locality, compared to naive broadcast, Nandi et al.'s algorithm makes sure that the computation for each batch area is done with local copy-add operations. However, the messages sent from input rows to the batch areas are remote, which can be costly.

Regarding balance, to handle cases where certain cube regions are reducer-unfriendly in terms of reducers' input sizes, which can cause reducers to run for a long time or simply fail, a separate sampling stage is used to estimate which cube regions are reducer-unfriendly, and ``value-partition" each such cube region into multiple ones along a big dimension, one for each value of the dimension, with a later MapReduce phase for combining the results.

Overall, besides being complex, Nandi et al.'s approach can send a high number of remote messages from every input row, leading to inefficiency. Our algorithm will send a small amount of remote messages, which turns out to be just a couple for each input row in average for our dataset.

\section{Our Algorithm}

We take a different approach to design an algorithm that is good in message minimization, locality, and balance.

A data segment $A$ is a primary child of a data segment $B$, if $A$ and $B$'s keys differ exactly at the position where $B$ has its rightmost * in $G$, where $A$ takes a concrete value. For example, (``US", ``CA", *, ``Retail", ``Amazon'') is a primary child of (``US", ``CA", *, *, ``Amazon'').
A simple observation from Gray et al.~\cite{G+97} is that the aggregate for $B$ can be computed directly from the counts of all its primary children. Given that we need to compute counts for all its children anyways as part of the materialization requirement, this is more efficient than aggregating over all input rows that belong to segment $B$, as in the broadcast algorithm.

\subsection{A Naive Algorithm}

We can layer the work of aggregating from primary children. If we have the counts for the set of all segments with $k$ aggregated columns, then we can determine the counts for the set of all segments with $k+1$ aggregated columns, because the primary children of the latter set all fall in the former set. This can be implemented using a chain of MapReduces with segment keys as the MapReduce keys to randomly shard by.

This algorithm is nearly optimal in message minimization. It is also excellent in balance as it shards by all but one columns. However, as the key sharding is random, most of the messages will be remote instead of local, which is undesirable, not to mention the setup cost of having a large number of MapReduce phases in practice.

\subsection{A Batched Algorithm}

To reduce the amount of remote messages, we use the idea of column batching which allows us to leverage locality. (This algorithm structures the aggregation differently, and will not be a generalization of the naive algorithm.)

Our algorithm takes as additional input a grouping $G_g$, $G_{g-1}$, .., $G_1$ of all columns into $g$ non-empty groups, where if we enumerate the columns from $G_g,..,G_1$ in order, it matches with the original column ordering. We will discuss the choice of grouping later.

We say a segment is concrete in $G_i$ if the values of all its columns in $G_i$ are not aggregated, or it is aggregated in $G_i$ otherwise.
Our algorithm consists of $g$ MapReduces, where it processes groups $1,..,g$ in sequence.

\begin{algorithm}
  \KwIn{rows with hierarchical dimensions and counts}
  \KwOut{counts for all segments}
  \KwSty{Additional input}: grouping $G_g,..,G_1$\\
\For{phase $i$ from $1$ to $g$}{
  Run a MapReduce with the following specs:\\
  \KwSty{Input for phase $i$}: all segments that are concrete in $G_g,G_{g-1},..,G_i$, and either concrete or aggregated in $G_{i-1},..,G_1$, with the counts\\
  \KwSty{Mapper for phase $i$}: work partitioning\\
  \KwSty{Reducer for phase $i$}: local materialization\\
  \KwSty{Output for phase $i$}: all segments that are concrete in $G_g,G_{g-1},..,G_{i+1}$, and either concrete or aggregated in $G_i,..,G_1$, with the counts\\
}
\caption{Outline of Our Algorithm}
\end{algorithm}

Note that these MapReduces chain, with the algorithm's input and output being the input for phase 1 and the output for phase $g$ respectively.

We observe that the work of the MapReduce for phase $i$ can be partitioned based on the values of all groups but $G_i$. To be precise, let $S$ be all segments with some fixed concrete values for $G_g,..,G_{i+1}$, fixed concrete or aggregated values for $G_{i-1},..,G_1$, and varying values for $G_i$. Counts for segments in $S$ that are aggregated in $G_i$ can be computed from those for segments in $S$ that are concrete in $G_i$. Essentially, we have a materialization problem for segments in $S$, and we will solve this materialization problem locally, using local messages.

Therefore in the algorithm, we can have the mappers partition the data. Remote messages are needed here.
\begin{algorithm}
\caption{Mapper for phase $i$: partition work}
\KwIn{(segment, count)}
\KwOut{(key, value)}
key := segment's values for all groups but $i$\;
value := (segment's values for group $i$, count)\;
emit (key, value);  // a remote message  \\
\end{algorithm}

Each reducer is then left with a local materialization problem, where we run the layer-by-layer Naive algorithm.\footnote{In our algorithm, every aggregation for a segment from its primary children happens within one machine. If a dimension is extremely large in cardinality, an aggregation can take a very long time. For such datasets, we could change the algorithm slightly to use post-mapper combiners \cite{DG04} to avoid this pitfall. The trick is that in the mapper phase, we do not simply copy group $i$'s concrete vector. Instead we can have each of them send messages to its unique primary parent w.r.t.~group $i$, so that for that each such parent, its count is computed from data from mappers in different machines, where post-mapper combiners can help to parallelize the computation.}

\begin{algorithm}
\caption{Reducer for phase $i$: materialize locally}

\KwIn{a list of (concrete values for group $i$, count)}
\KwOut{a list of (concrete or aggregated values for group $i$, count)}
  let each $h[k]$ for $k$ from $0$ to $|G_i|$ be a hash map that maps a vector of values of group $i$ to a count, where the vector contains exactly $k$ stars\\
  \For{(v, count) in input}{
    insert (v, count) into $h_0$, adding the count onto the existing value if any;  // a local message\\
    write $h_0$ to output, adding back values for all groups but $G_i$\;
  }
  \For{$k$ from 1 to $|G_i|$} {
    // compute $h_k$ from $h_{k-1}$ \\
    \For{($v$, count) in $h$}{
      \For{values $u$ for $G_i$ that $v$ is a primary child of}{
        insert ($u$, count) into $h_k$, adding count onto the existing value if any;  // a local message\\
      }
    }
    write $h_k$ to output, adding back values for all groups but $G_i$\;
  }
\end{algorithm}
\subsection{Properties of the Algorithm}
Observe that each phase of the algorithm contains the input in the output (for the first phase, input rows with the same key are aggregated first), and hence the data size grows phase by phase. Most cube materialization involves a nontrivial blow-up in data cube size as discussed in Section~\ref{problem}. This also happens phase by phase, and we call the output to input ratio for each phase as the \textbf{phase blow-up ratio}. The last phase is very important as it has the most work, and its phase blow-up ratio is crucial to the performance of the algorithm.

We study the algorithm in terms of the three factors.

On message minimization, the count of a segment is always computed from its primary children, as opposed to the input rows as in naive broadcast. In worst case, note that each segment can be the primary child of at most $n$ segments, where $n$ is the number of dimensions. In practice, this number is much smaller in average (which turns out to be less than 3 times the number of indistinct segments for our dataset).

On locality, for the last phase, where most work is done, the mapper uses one remote message to send each input row to the corresponding reducer, and all the remaining messages are local within reducers. If the phase blow-up ratio is large for the last phase, the number of output rows is much larger than the number of input rows, and it follows that local messages greatly outnumbers remote messages, implying good locality.

On balance, we partition work by the values of all but one groups. Since we we are sharding using many columns' values, this often results in granular sharding, which leads to balance. For our actual dataset, this flexibility helped us get around a complicated data-skewness with ease.

The performance of the algorithm is affected by the choice of the grouping. We should arrange the table schema and grouping so that columns with smaller cardinalities are in groups with smaller indices, to reduce the average number of primary children. We should form only two or three groups to avoid MapReduce setup cost, each with nontrivial data so that sharding by all but one groups leads to balance. Subject to balance, we leave more columns to the last group, so that the last phase has a big phase blow-up ratio for better locality.

\section{Empirical Study}

In a practical and important use case, we analyze a large dataset to help understand Google's revenue changes. The main metric is revenue change, and we have a long list of dimensions and their columns for capturing most potential causes. We have three types of dimensions, ones related to users, such as region, query-category-id, ones related to websites which display google's content ads, such as website-id, its category, and ones related to advertisers, such as id and category. We have a total of 11 dimensions with 14 columns in total. Three of them have large cardinalities ranging from 1K to 1M, four of them have sizes at most 4. The others are in between. For confidentiality reasons, we can not reveal the full list of dimensions\footnote{We do not use any data related to individual users such as their queries or identity information for our analyses.}.

Our goal is to compute a change metric for each segment, and outputs all segments with significant changes that meet a threshold. Pruning is difficult as a low-column segment can have a big change even if its parent segments do not.

Our dataset is big with 24.9G rows and 1.25TB of input data. The materialized cube has a size of 1.75TB.

Our dataset exhibits strong skewness. If we shard solely based on e.g.~advertiser-id, there exist big advertisers each of which contributes to a nontrivial fraction of the dataset. The same is true about website-id, and essentially every single dimension we have.\footnote{This is different from the dataset studied in Nandi et al., which has the user-id dimension. User-id is a good sharder, as no single user can contribute a large amount of data.}

To apply our algorithm over this dataset, we simply form three column groups, for users, websites, advertisers dimensions separately. Effectively, at every step, we shard the data using two dimension column groups instead of one, and this leads to even sharding.

\subsection{Analysis of the Run}

We ran our algorithm with 400 machines over the big dataset. In Table 2, for each of the MapReduce phases, we list the number of input rows, the total and maximum number of output rows across different MapReduce keys, and the total and maximum number of local messages across MapReduce keys. We also list the blow-up ratio of materialization in each phase, and the ratio of local messages to remote messages.

In the first phase, the dataset actually shrunk due to redundancy in input rows. After that, cube materialization started to blow-up, and the number of segments increased by a factor of 2.9 and 6.6 in the next two phases. The final cube size is 1.75TB with 35G tuples. (In contrast, Nandi et al.'s dataset had an input size of 55GB and a data cube size of 216GB.)

We look at the algorithm's performance regarding the three important factors.

\textbf{Message minimization}: The total number of distinct segments is 35.0G. If we do not count the one remote message for each input row, which seems unavoidable, we only send 7.1G remote messages, and 58.3G local messages.

In comparison, if we apply Nandi et al.'s algorithm, we expect a significant number of remote messages to be sent for each input row in the first phase (28 for a dataset in Nandi et al.~that is smaller with fewer dimensions than ours).

\textbf{Locality}: Excluding the one message we send for each input row, $58.3/(58.3+7.1) = 89\%$ of the messages were local. The key here is that we leveraged the fact that the blow-up factor of cube materialization is high for the dataset. We put enough columns in the last group so that the last phase had a significant blow-up ($5.3G \to 35.0G$), and then majority of the messages happened within this group as local messages, which outnumbered the remote messages that happened before.

\textbf{Balance}: No single key's reduce was in charge of more than 0.2\% of the local messages, or the nodes. Since all these reduce operations are randomly distributed over 400 machines, this led to good balance.

\begin{table}[hb]
\centering
\caption{Run Stats}
\begin{tabular}{|c|c|c|c|c|c|c|c|} \hline
phase & \#input & \#remote & \#output  & \#local & run \\
& rows & msgs & rows &                  msgs          & time\\ \hline
1 & 24.9G & 24.9G & 1.8G & 4.5G & 13min \\ \hline
2 & 1.8G & 1.8G & 5.3G & 8.0G  & 10min\\ \hline
3 & 5.3G & 5.3G & 35.0G & 45.6G & 31min \\ \hline
total & 32.0G & 32.0G & 42.1G & 58.3G & 54min\\ \hline
\end{tabular}
\begin{tabular}{|c|c|c|c|c|c|c|c|c|} \hline
phase & phase & \#local msgs/ & max \#nodes & max \#local\\
& blow-up & \#remote msgs  &  per key &  msgs per key \\ \hline
1 & & & 0.9M & 4.8M\\ \hline
2 & 2.9 &  4.4  & 1.5M & 3.0M\\ \hline
3 & 6.6 & 8.6 & 0.9M & 2.7M \\ \hline
\end{tabular}

\end{table}

\bibliographystyle{IEEEtran}

\bibliography{cube} 

\balance

\end{document}